\documentclass{b98proc}

\def\Journal#1#2#3#4{{#1} {\bf #2}, #3 (#4)} 

\def\NPA{{\em Nucl. Phys. } A}

\def\PLB{{\em Phys. Lett. } B}

\def\PRL{\em Phys. Rev. Lett.}
\def\PREV{\em Phys. Rev.} 

\def\PRD{{\em Phys. Rev.} D}
\def\PRC{{\em Phys. Rev.} C}
\def\ZPC{{\em Z. Phys.} C}
\def\ZPA{{\em Z. Phys.} A}

\def\shiftleft#1{#1\llap{#1\hskip 0.04em}}
\def\shiftdown#1{#1\llap{\lower.04ex\hbox{#1}}}
\def\thick#1{\shiftdown{\shiftleft{#1}}}
\def\b#1{\thick{\hbox{$#1$}}}

\begin{document}

\title{$N \to \Delta$ QUADRUPOLE TRANSITION IN THE CONSTITUENT QUARK MODEL}

\author{A. J. Buchmann\thanks{{\it E-mail address:} 
alfons.buchmann@uni-tuebingen.de}} 
\address{Institute for Theoretical Physics, University of T\"ubingen, \\ 
Auf der Morgenstelle 14, D-72076 T\"ubingen, Germany}

\maketitle

\abstracts{
Information on the intrinsic
deformation of the proton can be obtained by studying  the electromagnetic
$p \to \Delta^+$ quadrupole transition. Recent experiments have
shown that the electric quadrupole $(E2)$ strength in $\gamma p \to \Delta^+$ 
is about 10 times larger than predicted by the simple quark model 
using only one-body currents. Our analysis provides evidence 
for the dominance of exchange currents in the $N \to \Delta$ quadrupole 
transition, and  identifies the physical mechanism leading to the observed 
$E2$ strength.} 

\section{Introduction}

The quadrupole moment of a particle measures the deviation of its 
internal charge distribution from spherical symmetry. 
However, as a particle with total angular momentum $J=1/2$, 
the nucleon does not have a quadrupole moment in the laboratory frame.
In order to learn something about the shape 
of the nucleon one has to electromagnetically excite it, e.g.
to the $\Delta(1232)$ resonance, with total angular momentum $J=3/2$,
or to higher resonances.

There are three different ways to electromagnetically produce a
$\Delta(1232)$. Aside from the dominant 
$N \to \Delta$ magnetic dipole ($M1$) excitation mode, 
in which the spin of a single quark is flipped,
transverse electric quadrupole ($E2$), 
and longitudinal charge quadrupole ($C2$) transitions are allowed
by angular momentum and parity selection rules.
The strengths of these electromagnetic multipoles at 
photon three-momentum transfer ${\bf q}=0$, 
called $G_{M1}(0)$,  $G_{E2}(0)$, and $G_{C2}(0)$, can be 
extracted from high precision photo-pion- and electro-pionproduction 
experiments off the proton \cite{Bec97}. The quadrupole transition 
amplitudes  $G_{E2}(0)$ and $G_{C2}(0)$ are a measure of the intrinsic 
deformation of the nucleon.
The empirical $E2$ strength is with $G^{exp}_{E2}=0.133(20)$ \cite{Bec97}
about 10 times larger than predicted by the first quark model calculations 
using spatial single quark currents \cite{Isg82,Ger82,Dre84}. 

\section{Single Quark $N \to \Delta$ Quadrupole Transition }
\label{secstyle}

Early quark model calculations of the $N \to \Delta$ quadrupole transition 
used aside from a spin-independent confinement a
one-gluon exchange potential $V^{OGEP}$ between constituent quarks 
\cite{Isg82,Ger82}. Without retardation and for equal quark masses it reads
\begin{eqnarray}
\label{gluon}
V^{OGEP} ({\bf r}_i,{\bf r}_j) & = & {\alpha_{s}\over 4}
\b{\lambda}_i\cdot\b{\lambda}_j \Biggl \lbrace
{1\over r}-
{\pi\over m_q^2} \left ( 1+{2\over 3}
\b{\sigma}_{i}\cdot\b{\sigma}_{j} \right ) \delta({\bf r}) \nonumber \\
& & -{1\over 4m_q^2} {1\over r^3}
\left ( 3\b{\sigma}_i\cdot  {\hat {\bf r} } \,
\b{\sigma}_j\cdot{\hat{\bf r}}
-  \b{\sigma}_i\cdot\b{\sigma}_j \right ) \nonumber \\
&   & -{1\over 2m_q^2} {1\over r^3} \biggl [ 3 \left ( {\bf r}
\times {1\over2} ({\bf p}_i-{\bf p}_j) \right ) \cdot
{1\over2}(\b{\sigma}_i+\b{\sigma}_j) \nonumber \\
&  & - \left ( {\bf r}\times {1\over2} ({\bf p}_i+{\bf p}_j) \right ) \cdot
{1\over2}(\b{\sigma}_i-\b{\sigma}_j) \biggr ]
\Biggr \rbrace ,
\end{eqnarray}
where ${\bf r}={\bf r}_i-{\bf r}_j$;  $\b{\sigma}_i$ is the
Pauli spin matrix, and  $\b{\lambda}_i$ is the color operator of the i-th 
quark. Usually, a regularized 
one-pion exchange interaction between constituent quarks is introduced,
in order to account for the chiral symmetry of QCD and its
spontaneous breaking at low energies
\begin{eqnarray}
\label{pion}
V^{OPEP}({\bf r}_i,{\bf r}_j) 
& = &
{g_{\pi q}^2\over {4 \pi (4 m_q^2)}}
{1\over 3}{\Lambda^2\over {\Lambda^2-\mu^2}}
\b{\tau}_{i}\cdot \b{\tau}_{j}
\Biggl [ \b{\sigma}_{i}\cdot\b{\sigma}_{j}
\left ( \mu^2{e^{-\mu r}\over r}- 4 \pi \delta({\bf r} )  \right ) \nonumber \\
& & +\left( 3\b{\sigma}_{i}\cdot{\hat {\bf r} }\,
\b{\sigma}_{j}\cdot{\hat {\bf r} }
- \b{\sigma}_{i}\cdot\b{\sigma}_{j} \right )
\left ( 1+{3\over \mu r}  +{3\over (\mu r)^2} \right )
\mu^2{e^{-\mu r}\over r}  \nonumber \\
& & -(\mu \leftrightarrow \Lambda)  \Biggr ].
\end{eqnarray}
Here, $\mu$ is the
pion mass, $\Lambda$ the cut-off mass,
and $\b{\tau}_i$ denotes the isospin of the i-th quark.

The tensor terms in the effective gluon and pion exchange
potentials induce $D$-wave admixtures in the three-quark $N$ and $\Delta$ 
wave functions. For clarity we restrict ourselves to a two-state 
harmonic oscillator model for the wave functions:
\begin{eqnarray}
\label{wave}
\left \vert N \right \rangle & = &
a_S \left \vert (S=1/2 \,, L=0) J=1/2 \right \rangle +
a_D \left \vert (S=3/2 \,, L=2) J=1/2 \right \rangle  \nonumber \\
\left \vert \Delta \right \rangle & = &
b_S \left \vert (S=3/2 \,, L=0) J=3/2 \right \rangle +
b_D \left \vert (S=1/2 \,, L=2) J=3/2 \right \rangle,
\end{eqnarray}
where the $D$-states are of mixed symmetric type.
In Eq.(\ref{wave}), the inner spin $S$ couples with the
orbital angular momentum $L$ to the total angular momentum $J$ 
of the quarks. 
The $D$-wave probabilities $a_D^2$ and $b_D^2$ as calculated by different
authors \cite{Isg82,Ger82} lie between $0.2\%$ and $0.4\%$
(see table \ref{admix}). Ref.\cite{Dre84} obtains a larger $D$-state 
probability of $\sim 1\%$. In any case, the calculated 
$D$ state probabilities are much smaller than the $5\%$ $D$-state
admixture in the deuteron wave function. We will come back to this point.

\begin{table}[tb]
\caption[admixtures]{ 
{Admixture coefficients for the $S$ and mixed symmetric $D$ states
in the nucleon and $\Delta$ ground state wave 
function as calculated by different 
authors. }}
\begin{center}
\begin{tabular}[t]{| r | r | r | r | r |} \hline
 $N$ & $a_{S}$ & $a_{D}$  & $b_{S}$  & $b_{D}$    \\ \hline
Ref.\cite{Isg82}  & 0.93  & -0.04    & 0.97    & 0.07 \\  
Ref.\cite{Ger82}  & 0.95  & -0.04    & 0.97    & 0.07 \\  
Ref.\cite{Dre84}  & 0.94  & -0.09    & 0.96    & 0.10 \\  
Ref.\cite{Wey86}  & 0.906 &-0.045    & 0.994   & 0.056   \\
Ref.\cite{Buc91}  & 0.934 & -0.047   & 0.990   & 0.064    \\
\hline
\end{tabular}
\end{center}
\label{admix}
\end{table}
\begin{figure}[htb]
$$\mbox{
\epsfxsize 10.0 true cm
\epsfysize 7.0 true cm
\setbox0= \vbox{
\hbox { \centerline{
\epsfbox{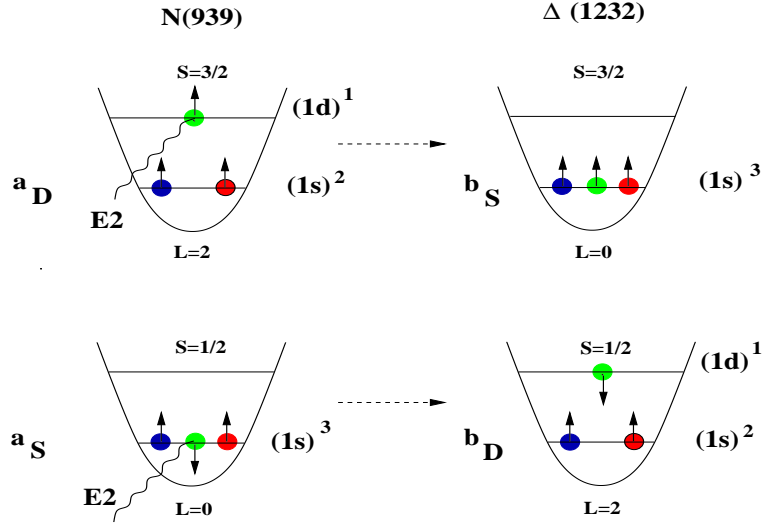}
}  } 
} 
\box0
} $$
\vspace{-0.5cm}
\caption[shell]{Conventional explanation
of the $\gamma N\to \Delta$ quadrupole transition via the one-body
current ${\bf J}_{[1]}$ in Fig.\ref{feynmec}(a)
(impulse approximation). In this approximation the
$E2$ transition amplitude is a coherent 
superposition of the two $L$ changing but $S$ conserving one-body 
transitions (upper and lower part of the figure).
In a single-quark transition, the absorption of a $E2$ photon is 
therefore only possible if either the nucleon (left) or the $\Delta$ (right) 
contains a $D$ wave admixture (deformed valence quark orbit). 
The transition probability amplitude 
$G_{E2}({\bf J}_{[1]})$ of Eq.(\ref{ge2imp}) 
is strongly suppressed due to the small $D$ wave admixtures 
in the $N$ und $\Delta$ wave function.}
\label{shellmodel}
\end{figure}

In the single quark transition model, based on the spatial one-body current
${\bf J}_{[1]}$, a quark moving in an excited $D$ state 
in the nucleon can absorb electromagnetic
quadrupole radiation ($L_{\gamma}$=2), and thus fall to the 
$S$ state in the $\Delta$. The amplitude for this process is 
proportional to $a_D b_S$. In addition, a quark in an 
$S$ state in the nucleon can absorb quadrupole radiation 
$(L_{\gamma}$=2), and jump to an excited $D$-state in 
the $\Delta$. This happens with an amplitude $a_Sb_D$. 
Both transitions, graphically displayed in Fig. \ref{shellmodel}, change 
the orbital angular momentum $L$, but 
leave the inner spin state $S$ of the quarks unchanged. 
The resulting single-quark $E2$ transition strength is \cite{Dre84} 
\begin{equation}
\label{ge2imp}
G_{E2}({\bf J}_{[1]})  =  {1 \over \sqrt{5}}
\left (a_{{D}} b_{{S}} + a_{{S}} b_{{D}} \right ).
\end{equation}
The $G_{E2}$ used here is dimensionless. 
The connection with the dimensionful transition quadrupole moment, 
or the helicity amplitudes is given in Ref. \cite{Buc97}. 
With the $D$ state admixtures calculated with the various potentials 
based on Eq.(\ref{gluon}) and Eq.(\ref{pion}) 
one obtains $G_{E2}=0.003-0.01$ \cite{Isg82,Ger82,Dre84}, 
i.e. theoretical results that differ by an order of magnitude from 
the recent experimental results. This suggests that some important 
dynamical feature is still missing.

\section{Exchange Currents and Siegert's Theorem}

From the theoretical point of view the conventional explanation 
outlined above is incomplete because it violates the continuity equation:
\begin{equation}
\label{totcc}
{\bf q }  \cdot {\bf J}^{}_{}({\bf q})= 
\left [ H, \rho_{}({\bf q}) \right ]. 
\end{equation}
After a decomposition of the charge and current operators 
into one- and two-body terms, i.e., 
${\rho}={\rho}_{[1]}+{\rho}_{[2]}$, and
${\bf J}={\bf J}_{[1]}+{\bf J}_{[2]}$,
one can show that the spatial two-body current ${\bf J}_{[2]}$ 
satisfies the following consistency relation
\begin{eqnarray}
\label{meccc}
{\bf q }\cdot {\bf J}_{[2]}({\bf q})
& = & \left [ V_{[2]}, \,  \rho_{[1]}({\bf q}) \right ]. 
\end{eqnarray}
\begin{figure}[t]
$$\mbox{
\epsfxsize 10.5 true cm
\epsfysize 3.0 true cm
\setbox0= \vbox{
\hbox { \centerline{
\epsfbox{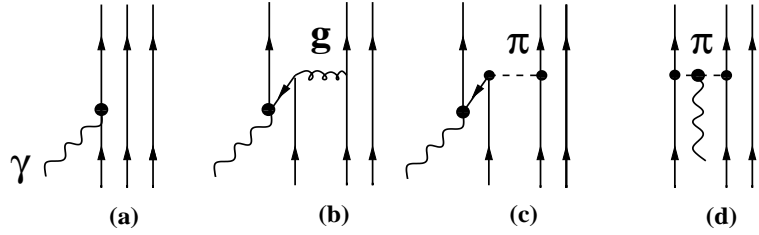}
}} 
} 
\box0
} $$
\vspace{-0.5cm}
\caption[Exchange currents]{Feynman diagrams of the four vector
current $J^{\mu}=(\rho, {\bf J})$:
(a) one-body current ${\bf J}^{\mu}_{[1]}$, 
and (b-d) two-body gluon and pion exchange currents 
${\bf J}^{\mu}_{[2]}$.
If the quarks interact via pion and gluon 
exchange, the two-body exchange currents depicted in  
diagrams (b-d) must be taken into account.
If only diagram (a) is considered, 
the continuity equation (\ref{totcc}) is violated.}
\label{feynmec}
\end{figure}
\noindent
Equation (\ref{meccc}) connects the quark-quark interactions
$V_{[2]}$, which determine 
the coefficients $a_S, a_D, b_S, b_D$ in Eq. (\ref{wave})
with the two-body currents ${\bf J}_{[2]}$
of Fig.(\ref{feynmec}), which determine the electromagnetic 
properties of the $N-\Delta$ system. Eq.(\ref{meccc}) is violated
if the potential contains momentum and/or isospin dependent terms
but the electromagnetic current contains only one-body terms.
In fact, the potentials of Eq.(\ref{gluon}) and Eq.(\ref{pion}) {\it do not} 
commute with the one-body charge operator, thus implying a two-body
current ${\bf J}_{[2]}$.

An important theorem based on Eq.(\ref{totcc}) is Siegert's 
theorem \cite{Sie37}. In the limit of small momentum transfers 
it relates the transverse electric $T^{E\,J}$ 
and longitudinal Coulomb multipoles $T^{C\,J}$:
\begin{equation}
\label{Sieg0}
\left \langle f \vert T^{E\,J}(\vert {\bf q}\vert \to 0 )
\vert i \right \rangle
= -{\omega \over \vert {\bf q} \vert} \, {\sqrt{{J+1} \over J}} \,
\left \langle f \vert T^{C\,J}(\vert {\bf q}\vert \to 0 )
\vert i \right \rangle.
\end{equation}
Thus, $G_{E2}$ can be calculated via the charge operator
(right hand side). The result will be the same as the one based on the 
spatial current operator (left-hand side). 

However, until recently, calculations of the $E2$ transition 
strength $G_{E2}$ using the spatial one-body current operator
have differed considerably (in some calculations by an order of magnitude) 
from those using the one-body charge operator
and Siegert's theorem. This important observation was made 
in Ref.\cite{Dre84}, and has been confirmed by other authors \cite{Bor87}.
Several improvements have been proposed, 
e.g. an increase of the number of harmonic oscillator states, 
and the inclusion of relativistic corrections to the one-body current, 
in order to remove this difference. However, the main reason for this 
discrepancy has not 
been recognized in previous works. We have recently shown \cite{Buc97}
that this difference is almost entirely explained by 
spatial two-body exchange currents ${\bf J}_{[2]}$ 
required by Eq.(\ref{meccc}) 
(see Table \ref{Siegtable}). 
Exchange currents were not explicitly included in previous analyses
of this problem.

\begin{table}[htb]
\caption[Sieg]{The transverse electric quadrupole form factor
$G_{E2}({\bf q}^2=0)$ for the $\gamma + p \to \Delta^+$ transition calculated
with (i) the one-body charge density $\rho^{}_{[1]}$
using Siegert's theorem (first row),
(ii) with the spatial current density ${\bf J}={\bf J}_{[1]}+{\bf J}_{[2]}$
(last row) for various quark models.
A comparison of the results in the first and last rows 
shows that the continuity equation is approximately satisfied, 
provided that the spatial two-body exchange currents ${\bf J}_{[2]}$ 
required by Eq.(\ref{meccc}) are included \cite{Buc97}. 
The remaining discrepancy between theory and recent experiment is 
removed by including $\rho_{[2]}$ (see Eq.(\ref{total})). }
\begin{center}
\nobreak
\begin{tabular}[htb]{| c | r | r | r | r | r |} \hline
 & Ref.\cite{Isg82} & Ref.\cite{Ger82} & Ref.\cite{Dre84} &
Ref.\cite{Wey86} & Ref.\cite{Buc91}($\pi$)
\\ \hline \hline
$G_{E2}(\rho^{}_{[1]})$       & 0.0192  & 0.0203   & 0.0796  & 0.0177
& 0.0165   \\   \hline
$G_{E2}({\bf J}^{}_{[1]})$    & 0.0118  & 0.0092  & 0.0076  & 0.0027
 & 0.0058    \\  \hline
$G_{E2}({\bf J}_{[2]})$        & 0.0084  & 0.0114   & 0.0561   & 0.0127
& 0.0105                                      \\ \hline
$G_{E2}({\bf J})$ & 0.0202  & 0.0206   & 0.0637
& 0.0154 & 0.0163  \\ \hline
\end{tabular}
\end{center}
\label{Siegtable}
\end{table}

\section{Double Spin Flip $N\to \Delta$ Quadrupole Transition}

As we have seen, a calculation using the one-body charge operator 
$\rho_{[1]}$ and Siegert's theorem yields substantially larger values 
for $G_{E2}$ as pointed out in Ref.\cite{Dre84}. We understand now why 
this is so. Most of the $\rho_{[1]}$ contribution to $G_{E2}$ comes from 
the two-body currents ${\bf J}_{[2]}$ (see Table \ref{Siegtable} and
Ref.\cite{Buc97} for further explanation). This finding suggests a 
different interpretation of the deformation of the nucleon in the quark 
model. However, there is still a large discrepancy with 
experiment when only $\rho_{[1]}$ is taken into account. We have shown that a 
double spin flip transition based on the two-body charge operator $\rho_{[2]}$
gives values for $G_{E2}$ in agreement with experiment \cite{Buc97}.
This is explained in the following.

The Coulomb quadrupole operator entering Eq.(\ref{Sieg0}) is:
\begin{equation}
\label{multipole}
T^{C2}(\vert {\bf q} \vert)=
         {1\over 4\pi} \int d\Omega_q \, \rho({\bf q}) \,
Y^{2}_0(\hat{\bf q}),
\end{equation}
where $\rho=\rho_{[1]}+\rho_{[2]}$ 
is the total charge operator. The quadrupole operator projects onto the 
$Y^{2}(\hat{\bf q})$, i.e., the quadrupole component in $\rho({\bf q})$. 
We discuss the quadrupole components in $\rho_{[1]}$ and $\rho_{[2]}$ 
separately. After a multipole expansion of the spin-independent one-body 
quark charge $\rho_{[1]}$ we see that the Coulomb quadrupole operator
is proportional to a second rank spherical harmonic
\begin{eqnarray}
\label{rho1}
T^{C2}(\rho_{[1]}) \propto Y^{2}({\hat{\bf r}_i }),
\end{eqnarray}
where ${\bf r}_i$ is a single-quark position coordinate. This is the only
allowed  tensor of rank 2 that can be constructed from a one-body
operator.
The operator in Eq.(\ref{rho1}) has nonvanishing matrix elements between 
the wave functions of Eq.(\ref{wave})
only for the off-diagonal $S \to D$, $ D \to S$, and
the diagonal $D \to D $ transitions.

\begin{figure}[t]
$$\mbox{
\epsfxsize 10.0 true cm
\epsfysize 4.5 true cm
\setbox0= \vbox{
\hbox{ 
\epsfbox{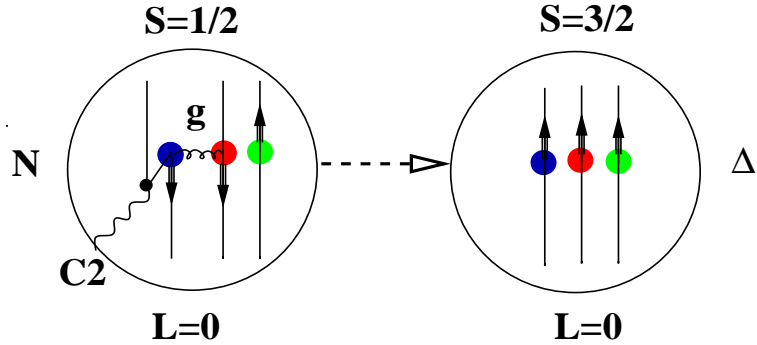}
} 
} 
\box0
} $$
\vspace{-0.8cm}
\caption[Spinflip]{Double spin flip due to, e.g., 
the {\it two-body} gluon 
exchange charge operator $\rho^{gq\bar q}_{[2]}$ of Fig.\ref{feynmec}(b).
The probability amplitude 
for this $S \to S$ transition is proportional to $a_S b_S \sim 1$, and thus 
considerably larger than the single-quark $E2$ transition amplitude 
in Eq.(\ref{ge2imp}), which is proportional to the small 
$D$ wave admixtures $a_D$ and $b_D$. 
A similar diagram can be drawn for 
the {\it two-body} pion 
exchange charge operator $\rho^{\pi q\bar q}_{[2]}$ of Fig.\ref{feynmec}(c).
The Coulomb quadrupole strength $G_{C2}$ and the related transverse 
electric quadrupole strength $G_{E2}$ calculated with the  
double spin flip is in good agreement with experiment 
\cite{Bec97}. }
\label{flips}
\end{figure}

On the other hand, the two-body gluon and pion exchange
charge densities $\rho_{[2]}$ contain, just like the corresponding  
potentials, a tensor in spin space, and the quadrupole operator is
\begin{equation}
\label{rank2}
T^{C2}(\rho_{[2]})  \propto 
[ \b{\sigma}^{1}_i \times  \b{\sigma}^{1}_j ]^{2}.
\end{equation}
Consequently, the operator in Eq.(\ref{rank2}) has a nonvanishing matrix 
element also for an
$S \to S$ transition. We stress that unlike the single-quark operator
in Eq.(\ref{rho1}), the two-body quadrupole operator in Eq.(\ref{rank2}) 
does not change the angular momentum of the wave function. 
However, as a tensor of rank 2 in spin space it simultaneneously 
flips the spin of two quarks.
The probability amplitude for this 
double spin flip transition is proportional to $a_S b_S \sim 1$, i.e.,
two orders of magnitude larger than the orbital angular momentum
changing one-body transition of Eq.(\ref{ge2imp}).
Morpurgo\cite{Mor89} has anticipated the important role 
of the operator in Eq.(\ref{rank2}) for the $\gamma p \to \Delta^+$ 
quadrupole transition.

The inclusion of two-body exchange currents leads to a
heretofore unknown relation between the mean square charge radii
of the $N$ and $\Delta$ \cite{Buc97}:
\begin{equation}
\label{radii}
r^2_p-r_{\Delta^+}^2=  r_n^2.
\end{equation}
Dillon and Morpurgo\cite{Mor99} recently derived Eq.(\ref{radii})
using a rather general QCD parametrization method 
and the assumption that three-index and loop terms are small.
They also estimate that if three-index and loop terms 
are included, deviations from Eq.(\ref{radii}) amount to $10-20\%$.

With the help of Eq.(\ref{radii}) one obtains a parameter-independent 
relation between the neutron charge radius $r_n^2$ and the $N \to \Delta$ 
transition quadrupole moment \cite{Buc97}    
\begin{eqnarray}
\label{relations}
Q_{p \to \Delta^+} & = & {1 \over \sqrt{2} }
(r^2_p-r_{\Delta^+}^2)={1 \over \sqrt{2} }  r_n^2.
\end{eqnarray}
Our parameter-independent prediction for transition quadrupole moment  
$Q^{theory}_{p \to \Delta^+}=-0.083$ fm$^2$ is in excellent 
agreement with the value $Q^{MAMI}_{p \to \Delta^+}=-0.086(13)$ fm$^2$
extracted from the MAMI data, and $Q^{LEGS}_{p \to \Delta^+}=-0.105(16)$ fm$^2$
extracted from the LEGS data. Furthermore, 
using the empirical charge radius of the neutron,
and Siegert's theorem, we obtain 
using both $D$ waves {\it and} exchange currents
\cite{Buc97}:
\begin{eqnarray}
\label{total} 
G_{E2}(0) & = & G_{E2}(\rho_{[1]})+ G_{E2}(\rho_{[2]})
\nonumber \\
& = & 0.017 + 0.107 =0.124.
\end{eqnarray}
This has to be compared to the experimental results which lie between 
$G^{exp}_{E2}=0.133(20)$ and $G^{exp}_{E2}=0.107(17)$ \cite{Bec97}. 
We also mention that the Z-diagrams in Fig.\ref{feynmec} not only explain 
$G_{E2}$ but also the experimental charge radius of the neutron \cite{Buc91}. 
This indicates that both the neutron charge radius and the $N \to \Delta$ 
quadrupole transition moment are dominated by nonvalence quark degrees
of freedom.

\section{Double spin flip in the deuteron}

The pion exchange induced double spin-flip term also contributes to the 
quadrupole moment of the deuteron. 
The quadrupole moment of the deuteron including the 
pion-pair current correction is given as 
\begin{eqnarray}
\label{deutquad}
Q_{d} & = &  {1\over \sqrt{50}} \int_0^{\infty} \!\!\!  dr r^2 \, u_2(r) 
\left ( u_0(r)-{1\over \sqrt{8}} u_2(r) \right ) \nonumber \\
& & + {f^2_{\pi N N } \over 4 \pi} {1 \over M_N} 
\int_0^{\infty} \!\!\!\! dr r \,  Y_1(\mu r)  
\left (2 u_0^2(r) -\sqrt{2} u_0(r) u_2(r) -{1\over 5} u_2^2(r) \right ).
\end{eqnarray}
The first term in (\ref{deutquad}) is due to the nonrelativistic 
one-body charge density $\rho_{[1]}$, while the
second term is due to the isoscalar two-body pion pair charge density
$\rho_{[2]}$,  and 
$Y_1(x)=(e^{-x}/x)(1+1/x)$.
Evidently, the one-body contribution is only nonzero if there is a 
nonzero $D$-state wave function $u_2(r)$. On the other hand, 
even for a pure (hypothetical) $S$-wave deuteron one obtains a 
nonvanishing quadrupole moment due to the pion-pair exchange current term.
The $u_0^2$ in the integrand is the contribution of the
double spin flip term. Numerically, the exchange current contribution to  
$Q_d$ is with 0.01 fm$^2$  rather small. The dominant
contribution of 0.28 fm$^2$ comes from the one-body charge density 
involving the $D$-state in the deuteron. 

Can one understand this role reversal
between $D$ states and exchange currents when going from a single baryon
to the deuteron? For a particle of mass $m$ moving in a 
harmonic oscillator potential there is an inverse proportionality between the 
excitation energy $\omega$, the size of the system $b$, and the mass $m$,
namely $\omega \propto 1/(mb^2)$. Because the average distance between two
quarks in the nucleon is approximately 1.0 fm, and the average distance 
between the nucleons in the deuteron is 4 fm, 
one needs $2 \hbar \omega \sim 600 $ MeV to lift a quark in an 
excited $D$ state, whereas one needs only $2 \hbar \omega \sim 4 $ MeV 
to lift a nucleon into a $D$ state. Given that the pion 
tensor potential is similar in strength in both systems, 
one can qualitatively understand 
why the $D$-state probability in the nucleon is so small compared to the one
in the deuteron.
On the other hand, the amount of particle-antiparticle pairs in a system 
is enhanced if the system size is decreased. Thus, the interchanged role of 
exchange currents and $D$-waves in the deuteron as compared to a single baryon
is mainly a consequence of the different size of these systems.

\section{Quadrupole Moment of the $\Delta$ }

Tensor forces and exchange currents 
also lead to a nonvanishing quadrupole moment of the $\Delta$. 
With configuration mixing but no exchange currents (impulse approximation)
one obtains neglecting the small $b_D^2$
contributions and with typical values for the
admixture coefficients \cite{Ger82} 
\begin{equation}
\label{c2imp}
Q^{imp}_{\Delta}= b^2 \, {4 \over \sqrt{30} }\left ( b_{S_S} b_{D_S}
+ {2\over \sqrt{3} } b_{S'_S} b_{D_S} \right ) e_{\Delta}=-0.087 b^2 
e_{\Delta}.
\end{equation}
For the quadrupole moment of the $\Delta$ the 
symmetric $D$ state amplitude $b_{D_S}$ is relevant. 
For $b=0.61$ fm one obtains
$ Q^{imp}_{\Delta}= -0.032 {\rm fm}^2 \, e_{\Delta}$.
Using $V^{OGEP}$ of Eq.(\ref{gluon}), Richard and Taxil, and 
Krivoruchenko and Giannini found
$Q^{imp}_{\Delta}=(2/5) r_n^2 e_{\Delta}$, which for $b=0.71$ fm 
coincides with Eq.(\ref{c2imp}).

On the other hand, with exchange currents but no configuration mixing 
we found  \cite{Buc97}
\begin{equation}
\label{deltaqm}
Q^{exc}_{\Delta}=r_n^2 e_{\Delta}. 
\end{equation}
Inserting the measured neutron charge radius into Eq.(\ref{deltaqm}) 
gives $ Q^{exc}_{\Delta}= -0.119 {\rm fm}^2 e_{\Delta}$.
Due to the smallness of the admixture coefficients in the $\Delta$ wave 
function, this result remains 
qualitatively unchanged if the two-body operator is
evaluated with mixed wave functions. 
Thus, nonvalence quark degrees of freedom 
effectively described here by quark-antiquark pair currents provide
the dominant contribution to the quadrupole moment of the $\Delta$.

\section{Deformation of the Nucleon}
 
This section is somewhat speculative. Because the nucleon has $J=1/2$, its
quadrupole moment in the laboratory frame is zero. This is analogous to 
a strongly deformed $J=0$ nucleus. All orientations of a deformed $J=0$ 
nucleus are equally probable. This results in a spherical charge distribution
in the ground state and a vanishing quadrupole moment $Q$ in the
laboratory. Nevertheless, one can obtain information 
on the {\it intrinsic} quadrupole moment\footnote{Here, {\it intrinsic} means 
with respect to a body-fixed coordinate system that rotates with the nucleus.}
$Q_0$ by measuring electromagnetic
quadrupole transitions between the ground and excited states,
or by measuring the quadrupole moment of an excited state 
with $J > 1/2$ of that nucleus.

In the collective model \cite{Boh75} the relation between the observable 
quadrupole moment $Q$ and the intrinsic quadrupole moment $Q_0$ is 
\begin{equation}
\label{collective}
Q= {3 K^2 -J(J+1) \over (J+1) (2J+3) } Q_0,
\end{equation}
where $J$ is the total spin of the nucleus,
and $K$ is the projection of $J$ onto the $z$-axis in the body fixed frame
(symmetry axis of the nucleus).
The intrinsic quadrupole moment $Q_0$ characterizes the deformation of the 
charge distribution in the ground state. 
A simple model for a nonspherical homogeneous charge distribution is that 
of a rotational ellipsoid with charge $Z$, major axis $a$ along, and minor 
axis $b$  perpendicular to the symmetry axis:
\begin{equation}
\label{ellipsoid}
Q_0= {2 Z \over 5} (a^2-b^2) ={4 \over 5} Z R^2 \delta,
\end{equation} 
with $\delta=2(a-b)/(a+b)$ and $R=(a+b)/2$.

Can one use this model to determine the sign and magnitude of 
the deformation of the nucleon? Because the quadrupole moment 
of the $\Delta$ and the $N \to \Delta$ transition quadrupole moment 
are dominated by gluon and pion degrees 
of freedom, and {\it not} by single quark degrees of freedom, 
the collective model maybe valid. 
Assuming that the collective model is applicable, we consider 
the $\Delta$ with spin $J=3/2$ as a collective rotation of the entire nucleon 
with an intrinsic angular momentum $K=1/2$. One then 
finds from Eq.(\ref{deltaqm})
and Eq.(\ref{collective}) $Q_0=0.565$ fm$^2$ for the intrinsic quadrupole 
moment of the nucleon. Furthermore, from the naive nucleon 
model of Eq.(\ref{ellipsoid}) one obtains with $R=\sqrt{5/3}r_p=1.113$ fm 
a deformation parameter $\delta \approx 0.57 $ 
and a ratio of major to minor semi-axes $a/b \approx 1.79$.
This magnitude of deformation seems to be quite large. Yet,
we speculate that the sign of the intrinsic quadrupole moment 
given by Eq.(\ref{collective}) is correct. If so, the nucleon
is a prolate spheroid. The $\Delta$ is clearly an oblate spheroid with 
an observable quadrupole moment given by Eq.(\ref{deltaqm}). 

Although the nucleon wave function is not observable,
one prefers to describe the deformation of the nucleon 
in terms of a $D$ state admixture in the quark wave function. 
If both $S$ and $D$ wave amplitudes have 
the same (opposite) sign, one obtains a positive (negative) intrinsic 
quadrupole moment, and a prolate (oblate) deformation, respectively. 
For example, the
sign of the relevant symmetric $D_S$ wave amplitude $b_{D_S}$ in  
the $\Delta$ wave function is opposite to the dominant $S$-wave 
amplitude $b_{S_S}$ (see Refs. \cite{Isg82,Ger82,Buc91}).
This leads to a negative quadrupole moment of the $\Delta^+$ and an oblate 
deformation of the $\Delta$, as can be seen from Eq.(\ref{c2imp}).  
Because of the negative $D$ wave amplitude in the nucleon wave 
function, one could conclude that the nucleon has an oblate deformation.
However, one would then miss the contribution of the nonvalence quark degrees 
of freedom. One can (via a unitary transformation) 
eliminate the two-body transition operator, as discussed 
by Friar \cite{Fri75} for the deuteron, and reexpress it in terms of a 
modified potential, or modified 
wave functions. The matrix element is thereby left invariant. 
If the nucleon has an intrinsic quadrupole moment $Q_0=0.565$ fm$^2$, the
nucleon wave function should have a large and positive $D$ wave 
amplitude, in contrast to the small and negative
$D$-state admixture obtained with the original potentials. 

\section{Summary} 

With the help of Siegert's theorem 
we have shown that the largest part of the $E2$ transition strength 
$G_{E2}$ in 
$\gamma N \to \Delta$ comes from a two-body spin flip due to exchange currents.
The effect of one-body currents on $G_{E2}$ is relatively 
small. Exchange currents also 
dominate the quadrupole moment of the $\Delta(1232)$ \cite{Buc97}, and 
are very important for the radiative $E2$ decays of all decuplet baryons 
\cite{Wag98}. The quadrupole moment of the $\Delta$ together 
with the $N\to \Delta$ transition quadruple moment can be seen as an indication
for an intrinsic nucleon deformation. 
We conclude that the intrinsic deformation of 
baryons lies mainly in the nonvalence 
quark degrees of freedom ($q\bar q$ pairs, pions, gluons).
Our prediction of the $E2$ transition amplitude, 
which is based on the parameter-independent relation of Eq.(\ref{relations}) 
and Siegert's theorem, is in very good agreement with experiment. 

\vskip 0.2 cm
\noindent
Acknowledgement: I thank H. Clement, D. Drechsel, and A. Faessler for 
stimulating discussions.

\end{document}